\documentclass[dvipsnames,format=sigconf]{acmart}

\usepackage{algorithm}
\usepackage{algorithmicx}
\usepackage{amsmath}
\algnewcommand{\LeftComment}[1]{\Statex \(\triangleright\) #1}
\usepackage{algpseudocode}

\newcommand{\etal}{{\textit{et al.}}}

\AtBeginDocument{%
  }

\newcommand{\eg}{\emph{e.g.,}}

\begin{document}

\title{Neuroevolution Neural Architecture Search for Evolving RNNs in Stock Return Prediction and Portfolio Trading}


\author{Zimeng Lyu}
\affiliation{%
  \institution{Rochester Institute of Technology}
  \city{Rochester}
  \country{USA}}
\email{zimenglyu@mail.rit.edu}

\author{Amulya Saxena}
\affiliation{%
  \institution{Rochester Institute of Technology}
  \city{Rochester}
  \country{USA}
}
\email{as7108@rit.edu}

\author{Rohaan Nadeem}
\affiliation{%
  \institution{Rochester Institute of Technology}
  \city{Rochester}
  \country{USA}
}
\email{rn7823@rit.edu}

\author{Hao Zhang}
\affiliation{%
  \institution{Rochester Institute of Technology}
  \city{Rochester}
  \country{USA}
}
\email{hzhang@saunders.rit.edu}

\author{Travis Desell}
\affiliation{%
  \institution{Rochester Institute of Technology}
  \city{Rochester}
  \country{USA}
}
\email{tjdvse@rit.edu}

\renewcommand{\shortauthors}{Lyu et al.}

\begin{abstract}
Stock return forecasting is a major component of numerous finance applications. Predicted stock returns can be incorporated into portfolio trading algorithms to make informed buy or sell decisions which can optimize returns. In such portfolio trading applications, the predictive performance of a time series forecasting model is crucial. In this work, we propose the use of the Evolutionary eXploration of Augmenting Memory Models (EXAMM) algorithm to progressively evolve recurrent neural networks (RNNs) for stock return predictions. RNNs are evolved independently for each stocks and portfolio trading decisions are made based on the predicted stock returns. The portfolio used for testing consists of the 30 companies in the Dow-Jones Index (DJI) with each stock have the same weight. Results show that using these evolved RNNs and a simple daily long-short strategy can generate higher returns than both the DJI index and the S\&P 500 Index for both 2022 (bear market) and 2023 (bull market).
\end{abstract}



\maketitle

\section{Introduction}

In the field of finance and particularly in the context of time series analysis, predicting stock returns is generally considered more feasible and common than predicting absolute stock prices for the following reasons: first, the statistical properties such as the mean, variance, and autocorrelation of stock returns tend to be stationary over time, which makes them feasible for statistical and machine learning models~\cite{tsay2005analysis}. Second, stock return is a relative measure, therefore more meaningful for investors who are more interested in percentage changes in investments. Third, investors are generally more interested in returns than absolute stock prices because stock returns directly relate to portfolio performance and risk management~\cite{hull1997options}.

Stock return prediction is crucial for numerous finance applications. Predictive analytics in stock returns significantly aids in optimum asset allocation and risk-adjusted portfolio construction~\cite{cipiloglu2022portfolio}. Furthermore, it plays a crucial role in risk managements, where forecasts are integral for metrics such as Value at Risk (VaR) and Conditional Value at Risk (CVaR) to estimate potential portfolio losses under varying market conditions~\cite{stulz2008rethinking}. Additionally, algorithmic trading strategies leverage return predictions to execute high-volume, high-speed trading strategies\cite{bali2016empirical}. On a corporate scale, return forecasts influence strategic financial decision making including timing for market entries or exits, capital restructuring, and planning mergers or acquisitions~\cite{iqmal2020macroeconomic}. Moreover, it opens new perspective into making macroeconomic forecasting and analysis under the predicted market and economic conditions~\cite{ratanapakorn2007dynamic}. 

In portfolio algorithmic trading, predicted stock returns are crucial for making informed high-frequency buy or sell decisions to optimize returns. These applications can simply be divided into two principal components. The first component is the predictive model, which needs to be adapt at processing and analyzing large volumes of high-frequency stock data to forecast stock returns. The second component is the portfolio trading strategy, which utilizes the predicted returns of each stock within the portfolio to make trading decisions. The goal of this method is not only maximize profit but also meticulously manage risk. Therefore the predictive performance of time series forecasting models is crucial for such portfolio trading algorithms as the prediction performance directly affects the portfolio trading returns. 


Designing robust time series forecasting models using neuroevolution for each stock within a portfolio was investigated. The predictive performance of these models had a direct impact on the trading returns of the portfolio. Preliminary work explored potential of neuroevolution (NE) neural architecture search (NAS) to evolve recurrent neural networks (RNNs) using the Evolutionary eXploration of Augmenting Memory Models (EXAMM) algorithm~\cite{ororbia2019examm}, which is capable of generating reliable stock return predictions to enable increased profits through simple trading strategies. We constructed a portfolio consisting of equally weighted companies across the thirty companies listed on the Dow-Jones Industrial Index (DJI), and RNNs were evolved for each of the DJI companies individually for predicting returns. 
\section{Related Work}
\subsection{Stock Prediction}
Time series stock prediction employs a variety of methods ranging from traditional statistical models, such as autoregressive integrated moving average (ARIMA) \cite{reddy2019predicting} and autoregressive conditional heteroskedasticity (ARCH) \cite{sreelekshmy-2017}, to more advanced machine learning techniques, as the classical and traditional statistical models generally fail to capture the complex, non-liner dynamics in stock data. Those models often require data to be stationary and cannot handle sudden shifts in volatility~\cite{kumbure2022machine}. On the other hand, machine learning models offer several advantages that address these limitations. They are capable of capturing non-linear relationships, dynamically learning from non-stationary data. They excel in handling large volumes of noisy data and can detect complex patterns that are often missed by traditional models\cite{leippold2022machine}. Selvin \etal demonstrated even basic machine learning models outperform classical methods \cite{sreelekshmy-2017}. Recent research in deep learning, provides more reliable predictions in volatile market conditions after major disruptions like the COVID-19 pandemic~\cite{Ferdinand-2023}. 

RNNs are commonly used for time series predictions, such as stock price, stock or market trends, and stock returns. Smith~\etal~\cite{smith2024stock}, Pawal~\etal~\cite{pawar2019stock}, and Ghosh~\etal~\cite{ghosh2019stock} make stock price predictions with long-short term memory recurrent neural networks (LSTM-RNN). Stock market trends have also been be predicted using RNNs~\cite{zhao2021prediction} and LSTMs~\cite{yao2018high}~\cite{gao2018share}. Zhang~\etal have even incorporated deep belief networks (DBN) with LSTM networks to forecast stock price movement~\cite{zhang2021predicting}.

Alongside these researches, there has been growing interest in applying genetic and evolutionary algorithms to the field of stock market prediction. Genetic algorithms offer a different approach to model optimization, relying on principles of natural selection and genetics to evolve solutions over successive generations \cite{mccall2005genetic}. For instance, Chung and Shin \cite{chung2018genetic} introduced an efficient genetic algorithm-based approach for feature selection in stock market prediction, as have Xie \etal \cite{xie2023efficient}. Jozefowicz \etal conducted empirical studies demonstrating the effectiveness of genetic algorithms in optimizing the hyperparameters of deep neural networks for financial forecasting \cite{jozefowicz2015empirical}. Thakkar~\etal applied information fusion-based genetic algorithm for stock price and trend prediction~\cite{thakkar2022information}. Chen~\etal used combination of genetic algorithm feature selection with LSTM networks for stock predictions~\cite{chen2020stock}.

\subsection{Financial Decision Making}

Neuroevolution and neural architecture search are less commonly used for financial market tradings and financial decision making. Nadkarni~\etal combined neuroevolution and principal component analysis (PCA) to make trading decisions in financial market~\cite{nadkarni2018combining}. Consunji~\etal  designed a neuroevolution algorithm called EvoTrader that evolves neural networks to predict future market price movements for bitcoin trading~\cite{pellon2021evotrader}. Slimene~\etal used neuroevolution to evolve neural networks for early financial distress prediction of Tunisian companies~\cite{belhaj2019neuroevolution}.

The predictive ability of RNNs can also aid in portfolio trading and portfolio optimizations. Cao~\etal proposed online portfolio analysis for high frequency trading with RNNs~\cite{cao2023novel}. Ta~\etal  utilized LSTM networks to make portfolio optimization based stock movement predictions~\cite{ta2020portfolio}. Zhang~\etal proposed an optimal hedging strategy using LSTM-RNN~\cite{zhang2021option}. Ma~\etal, Cao~\etal, and Sen~\etal implemented portfolio optimization models using deep learning RNN and LSTM models~\cite{ma2021portfolio}\cite{cao2020delafo}\cite{sen2021stock}. Time series forecasting models are also used in other financial decision making applications, such as online fraud detection on transactions using RNNs~\cite{wang2017session}, risk management for automatic trading systems using LSTM~\cite{silva2020automated}, foreign exchange rate forecasting using RNNs~\cite{dautel2020forex}, and economic factors forecasting with RNNs~\cite{pahlawan2021stock}.




\section{Method}
\subsection{EXAMM}

In comparison with other work, this work utilizes the Evolutionary eXploration of Augmenting Memory Models (EXAMM) neuroevolution algorithm~\cite{ororbia2019examm}~to evolve RNNs to make stock return predictions. EXAMM evolves progressively larger RNNs through a series of mutation and crossover (reproduction) operations from minimal seed networks that only contain input to output connections. The population of RNNs is distributed across distinct islands where genomes in each island are evolved independently to preserve gene diversity, and periodically perform inter-island crossover for information exchange. New genomes are generated through 11 types of mutations and 2 types of crossovers~\cite{ororbia2019examm}. Weights of new generated genomes were inherited from parents to significantly reduce the amount of training needed~\cite{lyu2021experimental}. Islands are repopulated periodically to prevent the performance of islands become stagnant~\cite{lyu2021improving}.

Compared to other time series forecasting models, the uniqueness of EXAMM comes from the flexibility in evolving networks using node and edge level mutations along with inter- and intra-island for genome reproduction, and as well as its weight inheritance method which allows generated genomes to effectively inherit parents' weights to achieve good fitness without requiring a large number of training epochs~\cite{lyu2021experimental}. The combination of using memory cells, recurrent edge connections, and weight inheritance make it extremely well suited for generating RNNs for time series forecasting models. The flexibility of choosing different types of mutations and crossovers during the evolutionary process allows it to evolved RNNs that are specifically tailored for each stock, and has the potential to custom evolve RNNs for hundreds or even thousands of stocks without hand redesigning and tuning them.

\subsection{Portfolio Trading Strateges}
\subsubsection{\bf{Long-Only Strategy}}
\label{long-strategy}
A simple trading strategy is long-only strategy, where all stocks are bought with cash and no short selling is allowed. In the long-only strategy, a stock is purchased when the predicted stock return is positive, and that position is held until the predicted return for the stock becomes negative. Algorithm \ref{alg:long} shows the simple portfolio long-only strategy. At time $t$, first sell any previously bought stocks if their predicted return is negative to allocate capital. Then buy all stocks with predicted positive returns with the currently available capital. Since stocks have equal weight in the portfolio, we use an equal amount of capital to purchase each stock. At time $t$, the capital to invest in stock $i$ is the available cash amount at time $t$ divided by the total number of stocks that have positive predicted stock return: 
\begin{equation}
Capital[t,i] = CashAmount[t]/Count(StocktoInvest[t])
\end{equation}
At the end of the testing period, all holdings are liquidated and which is used to calculate the overall investment return. The portfolio trading return is calculated by:
\begin{equation}
Return = (C_1 - C_0) / C_0
\end{equation}
In the equation above, $C_0$ is the total initial capital for investments. At the end of the testing period, all stock holdings are cleared and $C_1$ is the overall cash collected at the end of the testing period.

The success of using long-only strategy is based on the expectation that the stock price of invested stocks should gradually increase, so that they will eventually generate positive investment returns. However the Long-only investment strategies inherently pick up market risk and are fully exposed to market fluctuations~\cite{curtis2004modern}. During a bull market, the long-only strategy typically outperforms the market with significant gains. However, during a bear market, such a strategy can lead to significant loss\cite{elbannan2015capital}.

\subsubsection{\bf{Daily Long-Short Strategy}}
The daily long-short trading strategy is more commonly used in finance literature~\cite{park2020intelligent, hou2020enriched}. Compared to a long-only strategy, a daily long-short strategy offers enhanced risk management by hedging against market volatility. This strategy is capable of generating returns in both bull and bear markets by taking long positions in stocks expected to appreciate in value and short positions in those expected to depreciate. 

It is commonly practice to allocate (or borrow) 50\% of the total investment portfolio to capital designated for short selling. For example, if the initial investment capital is \$100, an equivalent amount of \$100 can be allocated for short selling, culminating in a total investment portfolio valued at \$200. Therefore 50\% of the \$200 is allocated for short selling. Algorithm \ref{alg:long-short} demonstrates such simple long-short trading strategy. We first define the hyperparameters $NumStocksLong$ as the number of stocks to long, and $NumStocksShort$ as the number of stocks to short. 

At the beginning of time [t], the portfolio is sorted by predicted stock returns. Transactions are only executed at time [t] if all $NumStocksLong$ stocks demonstrate a positive predicted return, and all $NumStocksShort$ stocks demonstrate a negative predicted return. If those conditions are met, first all current positions are liquidated, which includes selling any held stocks and repaying shares that were borrowed for short selling. Based on the available capital at time [t], the amount of capital for taking long positions and short selling is then calculated.

\begin{algorithm*}[!t]
    \begin{algorithmic}[1]
        \caption{Long-only Strategy}
        \label{alg:long}
        \footnotesize
        \Function{Portfolio-Long-Only-Return}{PortfolioList, InitialCapital}
            \State $CurrentCash = InitialCapital$
            \Comment{\emph{With long-only, all stocks are bought with cash}}
            \For {$t$ in [1, T]}
                \For{$stock$ in $PortfolioList$}
                \Comment{\emph{At begin of time t, sell previously invested stocks if predicted return < 0}}
                    \If{$stock.predictedReturn(t) < 0$}
                            \State $CurrentCash += stock.sell(t)$
                            \Comment{\emph{sell(t) will return zero if hold no position at time t}}
                    \EndIf
                \EndFor
                \State $InvestList = []$ 
                \Comment{\emph{List of stocks to invest in at time t}}
                \For{$stock$ in $PortfolioList$}
                    \If{$stock.predictedReturn(t) > 0$} 
                    
                        \State $InvestList.add(stock)$
                        \Comment{\emph{Add the stock to list if predicted return is positive}}
                    \EndIf
                \EndFor
                \State $QuotaPerstock = CurrentCash / count(InvestList)$
                \Comment{The amount of money used for invest in each stock is equal}
                \For{$stock$ in $InvestList$}
                    \State $stock.buy(QuotaPerstock, t)$
                \EndFor
                \State $CurrentCash = 0$
            \EndFor
            \State $CurrentCash += ClearHoldings(PortfolioList)$ 
            \Comment{\emph{At the end of time T, clear all holdings in the portfolio}}
            \State $PorfolioReturn = (CurrentCash - InitialCapital/InitialCapital)$
            \Comment{\emph{Calculate overall return}}
        \EndFunction
    \end{algorithmic}
\end{algorithm*}

\begin{algorithm*}[!t]
    \begin{algorithmic}[1]
        \caption{Long-Short Strategy}
        \label{alg:long-short}
        \footnotesize
        \Function{Portfolio-Long-Short-Return}{PortfolioList, InitialCapital, NumStocksLong, NumStocksShort}
            \State $CurrentCash = InitialCapital$

            \For {$t$ in [1, T]]}
                \State $SortedPortfolio = Sort(PortfolioList, t)$ 
                \Comment{Sort portfolio by predicted stock return at time t, in descending order}
                \If{SortedPortfolio[NumStocksLong] > 0 and SortedPortfolio[-NumStocksShort] < 0}
                \Comment{Only trade at time t if conditions are met}
                    \State $CurrentCash += ClearHoldings(PortfolioList)$ 
                    \Comment{\emph{First clear all long and short holdings }}
                    \State $QuotaForLong = CurrentCash/NumStocksLong$
                    \State $QuotaForShort = CurrentCash/NumStocksShort$ 
                    \Comment{\emph{The capital allocated for short selling takes 50\% of total investment portfolio}}
                    
                    \For{$i$ in $[1, NumStocksLong]$}
                        \Comment{\emph{Long stocks}}
                        \State $SortedPortfolio[i].buy(QuotaForLong, t)$
                        \State $CurrentCash -= QuotaForLong$
                    \EndFor
    
                    \For{$i$ in $[1, NumStocksShort]$}
                        \Comment{\emph{Short stocks}}
                        \State $SortedPortfolio[-i].sell(QuotaForShort, t)$
                        \State $CurrentCash += QuotaForShort$
                    \EndFor

                \EndIf
            \EndFor
            \State $CurrentCash += ClearPortfolio(PortfolioList)$
            \Comment{\emph{At the end of time T, clear all holdings in the portfolio}}
            \State $PorfolioReturn = (CurrentCash - InitialCapital/InitialCapital)$
            \Comment{\emph{Calculate overall return}}
        \EndFunction

    \end{algorithmic}
\end{algorithm*}
\section{CRSP Dataset}

The Center for Research in Security Prices (CRSP) dataset is a comprehensive collection of data on the U.S. stock market. It contains all stocks on the NYSE, AMEX, and NASDAQ, and covers wide range of market indexes, open-end mutual funds, exchange-traded funds (ETFs), U.S. Treasury securities, and corporate bonds. CRSP data is known for its high level of accuracy and is highly reliable for academic research and financial modeling~\cite{crsp_2023,lemieux2014clustering}. CRSP data has a long historical range from December 1925 to present, with daily, monthly, quarterly, and annual data. The dataset contains valuable variables for financial research such as stock prices, returns, trading volumes, and dividend yields.

\subsection{Portfolio Selection}
In portfolio trading related literature, the portfolios used for testing generally contains hundreds or even thousands of stocks~\cite{almahdi2017adaptive}~\cite{park2020intelligent}, and the most commonly used portfolios include the S\&P 500 index and the Russell 1000 index. As this work focus on testing forecasting ability of EXAMM in predicting stock returns and maximizing returns on the simple portfolio trading strategies, we instead use the companies from a small yet well-known index -- Dow-Jones Industrial Average (DJI) as our portfolio.

In particular, we use the 30 Dow Jones Industrial Average (DJI) companies before the index change which occurred on Feb 26th 2024 as our portfolio. The current companies in the Dow Jones Industrial Average (DJI) include a diverse range of sectors from technology to healthcare. Some of the notable companies are the Microsoft Corporation, Apple Inc, JP Morgan Chase \& Co., Visa Inc., Walmart Inc., United Health Group Incorporated, The Procter \& Gamble Company, Johnson \& Johnson, and The Home Depot, Inc. Other companies in the index are Merck \& Co., Inc., Chevron Corporation, Salesforce, Inc., The Coca-Cola Company, The Walt Disney Company, Cisco Systems, Inc., McDonald's Corporation, Caterpillar Inc., Verizon Communications Inc., International Business Machines Corporation, American Express Company, Intel Corporation, Amgen Inc., NIKE, Inc., Walgreens Boots Alliance Inc., (WBA) Goldman Sachs Group, Inc., Honeywell International Inc., The Boeing Company, The Travelers Companies, Inc., 3M Company, and Dow Inc.

The DJI index is a price-weighted index, which means that companies with higher stock prices have more influence on the index's performance regardless of the actual size or market capitalization of the company. In our experiments, the portfolio is consisted of the 30 DJI companies and each company instead has equal weight in the portfolio, as we are investigating the overall performance of our methodologies.

\subsection{Predictors}
We choose 7 economic predictors that are associated with stock returns in stock data. Those predictors are directly obtained from the CRSP dataset or calculated by using variables form CRSP dataset. The 7 predictors are shown in table~\ref{table:financial_metrics}. 

  

\begin{table}[h]
\caption{Economic Predictors for Stock Return Prediction}
\label{table:financial_metrics}
\centering
\begin{tabular}{ll}
\hline
\textbf{Predictors} & \textbf{Description} \\
\hline
\emph{\textbf{Return}} & Percentage change in stock price \\
\emph{\textbf{Volume Change}} & Percentage change in trading volume \\
\emph{\textbf{Bid-Ask Spread}} & \((AskPrice-BidPrice)/Price\) \\
\emph{\textbf{Illiquidity}} & \(Return / (Volume \times Price)\) \\
\emph{\textbf{Turn Over}} & \(Volume / SharesOutstanding\) \\
\emph{\textbf{DJI Return}} & DJI index return \\
\emph{\textbf{S\&P 500 Return}} & S\&P 500 index return \\
\hline
\end{tabular}
\end{table}

\emph{Return} is the percentage change of stock price from time [$t-1$] to time [$t$], which in the case of this data is day-to-day, it is also the forecasting output parameter. \emph{Volume Change} is the percentage change of stock trading volume from time [$t-1$] to time [$t$]. \emph{Bid-Ask Spread} measures the difference between the highest price buyers are willing to pay (\emph{Bid}) and the lowest price the sellers are willing to sell (\emph{Ask}), and the \emph{Ask} price is usually higher than the \emph{Bid} price. \emph{Illiquidity} measures if the stock can easily be sold for cash. \emph{Turn Over} measures the rate stocks being sold at given period, and it is also a measure of stock liquidity. \emph{DJI Return} and \emph{S\&P 500 Return} are the DJI and S\&P 500 index returns, and they often serve as benchmarks for market performance. The predictors contain information for stock performance as well as market-wide information, and have association with stock returns \cite{foucault2013market, malinova2008liquidity, grimes1977stock}.

\subsection{Datasets}
The length of historical data for each of the DJI companies are different, but more than half of the 30 DJI companies' historical stock data are available in CRSP dataset since late 1992. We use the year of 1992 as the starting year of the training data. As newer companies such as Amazon Inc. and VISA Inc. do not have historical stock data going back to the late 1990s or 2000s, we then use the stock data from whenever it became available.

The actual market return on DJI index is $-8.78\%$ in the year of 2022 and $13.7\%$ in the year of 2023, which provide a good avenue for into the stock return performance and its affect in portfolio trading returns for both bull and bear years. We created two datasets to test the performance of stock return predictions. The first dataset contains stock data from 1992 to 2020 as training data, stock data in year 2021 as validation data, and stock data in 2022 as testing data. The second dataset contains training data from 1992 to 2021, validation data from year of 2022, and testing data from year of 2023. For simplicity, we name the two datasets by the year of the testing data. The first dataset is called the 2022 DJI company dataset, and the second dataset is called the 2023 DJI company dataset. 

For both datasets, the validation and testing data are the stock data of an entire calendar year, which is 250 to 252 data points. The length of training data varies among different companies. For example, in the 2022 DJI company dataset, Walgreens Boots Alliance Inc. has 1511 rows of training data, and Apple Inc has 7306 rows of training data. 
\section{Results}

\subsection{Experiment Setup}

\begin{table} [!t]\centering
\caption{Portfolio Trading Returns using Long-only Strategy Incorporated with Different Transaction Costs}\label{tab:return_tc}
\begin{tabular}{c|ccc}\toprule
     &  RET (\%) & RET w/ TC 1 (\%) & RET w/ TC 2 (\%) \\
\midrule
2022 & {39.52} & {38.78} & {38.94} \\
2023 & {-9.44} &{-9.62} & {-9.42} \\

\bottomrule
\end{tabular}
\end{table}


\begin{table} [!t]\centering
\caption{Portfolio Trading Returns Using Different Forecasting Models using Long-only Strategy}\label{tab:return_all}
\begin{tabular}{c|cc}\toprule
     &  2022 &  2023 \\
\midrule
EXAMM & -9.44 & \bf{39.05} \\
\midrule
LSTM & -11.00 &  8.10  \\
MGU & -16.01 &  16.82  \\
GRU & \bf{-8.38} &  14.94  \\
\midrule
AR & -16.16 & 12.66\\
ARIMA & -21.24  & 22.42 \\
SVR & -25.18  &  12.74 \\
RF & -18.01   & 15.01  \\

\bottomrule
\end{tabular}
\end{table}

\begin{table}[!htp]\centering
\caption{Buy \& Hold and Market Index Return}\label{tab:market}
\begin{tabular}{c|cccc}\toprule
 & Buy \& Hold & DJI Index & S\&P 500 \\\midrule
2022  &-1.60  &-8.78 &-18.11 \\
2023  &3.23 &13.70  &26.29 \\
\bottomrule
\end{tabular}
\end{table}

Each EXAMM run for each stock was repeated $10$ times using [REMOVED FOR REVIEW]'s research computing systems. This system consists of 2304 Intel® Xeon® Gold 6150 CPU 2.70GHz cores and 24 TB RAM, with compute nodes running the RedHat Enterprise Linux 7 system. Each EXAMM run utilized 16 cores.

\begin{table*}[!htp]\centering
\caption{2022 Daily Long-Short Strategy Portfolio Return}\label{tab:2022_Hybrid}
\scriptsize
\begin{tabular}{l|cccccccccccccccc}\toprule
&short\_1 &short\_2 &short\_3 &short\_4 &short\_5 &short\_6 &short\_7 &short\_8 &short\_9 &short\_10 &short\_11 &short\_12 &short\_13 &short\_14 &short\_15 \\\midrule
long\_1 & -15.63 & -25.12 & -29.44 & -18.60 & {\bf{-3.85}} & {\bf{0.42}} & {\bf{6.14}} & {\bf{3.44}} & {\bf{-1.41}} & {\bf{1.05}} & {\bf{0.85}} & {\bf{-6.73}} & {\bf{2.85}} & {\bf{4.47}} & {\bf{4.81}} \\
long\_2 & -15.10 & -21.36 & -22.95 & -21.14 & -18.8 & -14.51 & {\bf{-7.94}} & -9.16 & -13.40 & -12.33 & -10.97 & -13.44 & {\bf{-4.75}} & {\bf{0.77}} & {\bf{-3.13}} \\
long\_3 & -13.87 & -15.49 & -14.34 & -18.12 & -15.59 & -12.95 & -11.43 & -11.42 & -11.53 & -10.30 & {\bf{-6.41}} & {\bf{-1.18}} & {\bf{6.27}} & {\bf{10.00}} & {\bf{5.00}} \\
long\_4 & {\bf{-5.06}} & {\bf{-7.75}} & -8.71 & {\bf{-5.45}} & {\bf{1.33}} & {\bf{0.55}} & {\bf{-1.76}} & {\bf{-3.61}} & {\bf{-0.66}} & {\bf{2.15}} & {\bf{4.31}} & {\bf{6.12}} & {\bf{12.09}} & {\bf{13.59}} & {\bf{6.09}} \\
long\_5 & {\bf{-8.00}} & -10.05 & -13.47 & -11.96 & {\bf{-6.64}} & -11.68 & -13.12 & -11.55 & -9.18 & {\bf{-2.45}} & {\bf{1.76}} & {\bf{5.76}} & {\bf{12.84}} & {\bf{12.85}} & {\bf{8.63}} \\
long\_6 & -14.96 & -17.37 & -22.14 & -22.41 & -17.57 & -17.98 & -19.04 & -19.64 & -17.12 & {\bf{-8.44}} & {\bf{-6.31}} & {\bf{-0.78}} & {\bf{6.25}} & {\bf{5.39}} & {\bf{3.50}} \\
long\_7 & -11.23 & -17.61 & -23.47 & -23.26 & -18.68 & -17.68 & -18.45 & -18.17 & -16.79 & -10.06 & -10.48 & {\bf{-7.53}} & {\bf{2.05}} & {\bf{1.50}} & {\bf{1.24}} \\
long\_8 &-9.12 &-18.83 &-26.27 &-23.58 &-19.5 &-19.42 &-20.13 &-19.96 &-21.01 &-14.94 &-14.09 & {\bf{-7.98}} &{\bf{-1.06}} &{\bf{-0.66}} &{\bf{-0.96}} \\
long\_9 &{\bf{-8.76}} &-15.22 &-22.58 &-17.42 &-14.56 &-14.75 &-14.16 &-13.59 &-16.39 &-11.45 &{\bf{-0.11}} &{\bf{11.42}} &{\bf{11.04}} &{\bf{11.49}} &{\bf{8.68}} \\
long\_10 &{\bf{-2.40}} &-14.20 &-16.82 &-10.71 &-9.15 &-9.89 &{\bf{-7.81}} &{\bf{-8.76}} &{\bf{-6.72}} &{\bf{-1.83}} &{\bf{11.30}} &{\bf{17.37}} &{\bf{18.57}} &{\bf{18.30}} &{\bf{14.26}} \\
long\_11 &{\bf{0.39}} &{\bf{-8.05}} &-15.61 &-19.20 &-13.66 &-14.14 &-11.95 &-15.15 &-12.33 &{\bf{8.86}} &{\bf{19.30}} &{\bf{24.64}} &{\bf{28.01}} &{\bf{27.52}} &{\bf{20.14}} \\
long\_12 &{\bf{-8.74}} &-21.03 &-24.06 &-22.38 &-12.48 &-11.75 &-14.53 &-15.99 &-14.40 &{\bf{5.92}} &{\bf{10.89}} &{\bf{23.29}} &{\bf{25.67}} &{\bf{21.97}} &{\bf{19.08}} \\
long\_13 &{\bf{-3.95}} &-18.56 &-20.69 &-19.35 &-12.26 &-9.37 &-15.23 &-14.96 &-13.77 &{\bf{-5.99}} &{\bf{-7.61}} &{\bf{6.03}} &{\bf{8.12}} &{\bf{10.92}} &{\bf{7.89}} \\
long\_14 &{\bf{-8.64}} &-25.07 &-22.65 &-19.56 &-14.76 &-10.90 &-17.02 &-15.4 &-13.74 &{\bf{-2.88}} &{\bf{-8.15}} &{\bf{5.26}} &{\bf{4.77}} &{\bf{7.81}} &{\bf{2.68}} \\
long\_15 &-13.91 &-27.81 &-27.49 &-19.43 &-15.93 &-13.12 &-17.77 &-21.05 &-19.71 &-12.73 &-15.70 &{\bf{-3.29}} &{\bf{-4.82}} &{\bf{-1.95}} &{\bf{-2.60}} \\

\bottomrule
\end{tabular}
\end{table*}

\begin{table*}[!htp]\centering
\caption{2023 Daily Long-Short Strategy Portfolio Return}\label{tab:2023_hybrid}
\scriptsize
\begin{tabular}{l|cccccccccccccccc}\toprule
&short\_1 &short\_2 &short\_3 &short\_4 &short\_5 &short\_6 &short\_7 &short\_8 &short\_9 &short\_10 &short\_11 &short\_12 &short\_13 &short\_14 &short\_15 \\\midrule
long\_1 &-1.48 &{\bf{14.55}} &{\bf{20.13}} &{\bf{27.79}} &10.33 &{\bf{17.84}} &12.66 &10.59 &13.02 &12.50 &{\bf{15.17}} &{\bf{27.74}} &{\bf{28.53}} &{\bf{24.03}} &{\bf{26.64}} \\
long\_2 &-1.24 &{\bf{18.76}} &{\bf{25.81}} &{\bf{30.92}} &{\bf{27.39}} &{\bf{26.68}} &{\bf{29.50}} &{\bf{19.68}} &{\bf{22.96}} &{\bf{17.95}} &{\bf{17.16}} &{\bf{22.73}} &{\bf{19.74}} &{\bf{16.19}} &13.22 \\
long\_3 &-4.81 &{\bf{15.84}} &{\bf{25.53}} &{\bf{27.45}} &{\bf{17.89}} &{\bf{17.61}} &{\bf{20.71}} &12.26 &13.18 &6.36 &-3.08 &1.88 &-1.71 &-4.08 &-4.42 \\
long\_4 &-11.93 &7.33 &{\bf{19.19}} &{\bf{25.61}} &{\bf{22.04}} &{\bf{20.55}} &{\bf{22.92}} &{\bf{16.08}} &{\bf{17.39}} &12.76 &6.98 &12.52 &7.75 &5.05 &3.18 \\
long\_5 &-20.54 &-1.18 &7.87 &{\bf{14.76}} &{\bf{14.53}} &13.03 &{\bf{14.96}} &9.71 &11.01 &11.11 &11.58 &{\bf{16.13}} &11.69 &11 &7.51 \\
long\_6 &-16.67 &0.91 &10.59 &{\bf{21.57}} &{\bf{23.76}} &{\bf{16.08}} &{\bf{18.03}} &{\bf{15.58}} &{\bf{18.11}} &{\bf{18.09}} &{\bf{17.67}} &{\bf{19.86}} &{\bf{15.54}} &{\bf{14.95}} &11.61 \\
long\_7 &-14.59 &3.74 &13.28 &{\bf{20.82}} &{\bf{22.27}} &{\bf{16.73}} &{\bf{18.98}} &{\bf{15.52}} &{\bf{19.28}} &{\bf{18.13}} &{\bf{19.10}} &{\bf{20.62}} &{\bf{16.44}} &{\bf{15.65}} &12.47 \\
long\_8 &-18.57 &1.10 &7.79 &{\bf{16.75}} &{\bf{20.25}} &13.05 &{\bf{16.54}} &11.87 &{\bf{17.56}} &{\bf{17.63}} &{\bf{21.40}} &{\bf{23.06}} &{\bf{19.54}} &{\bf{18.82}} &{\bf{15.03}} \\
long\_9 &-22.07 &-2.70 &4.97 &{\bf{16.32}} &{\bf{19.31}} &{\bf{15.26}} &{\bf{17.12}} &12.03 &{\bf{16.40}} &{\bf{14.60}} &{\bf{17.60}} &{\bf{18.32}} &{\bf{14.38}} &12.65 &8.47 \\
long\_10 &-25.82 &-3.84 &5.75 &{\bf{15.81}} &{\bf{19.31}} &{\bf{17.91}} &{\bf{18.61}} &{\bf{16.01}} &{\bf{15.34}} &{\bf{14.66}} &{\bf{16.35}} &{\bf{13.82}} &8.68 &7.13 &2.17 \\
long\_11 &-23.02 &-0.08 &7.18 &{\bf{14.28}} &{\bf{18.37}} &{\bf{16.09}} &{\bf{15.98}} &12.63 &10.16 &10.67 &11.57 &5.32 &0.97 &0.13 &-4.70 \\
long\_12 &-15.34 &3.65 &9.63 &7.64 &13.25 &7.43 &6.97 &5.60 &4.95 &5.15 &11.11 &5.26 &0.25 &-0.22 &-3.97 \\
long\_13 &-14.26 &4.95 &{\bf{14.22}} &10.85 &{\bf{15.18}} &11.08 &12.60 &7.57 &7.01 &9.41 &11.03 &7.96 &2.33 &-2.97 &-10.39 \\
long\_14 &-15.21 &1.46 &{\bf{14.11}} &10.15 &{\bf{15.66}} &10.73 &13.38 &7.76 &11.54 &12.08 &{\bf{14.01}} &12.08 &9.27 &2.67 &-5.16 \\
long\_15 &-9.85 &11.87 &{\bf{18.35}} &5.56 &13.47 &9.07 &11.5 &9.68 &{\bf{14.64}} &{\bf{13.76}} &11.17 &7.97 &7.22 &2.68 &-5.98 \\
\bottomrule
\end{tabular}
\end{table*}

Each EXAMM run used $10$ islands, each with a maximum capacity of $10$ genomes. Recurrent connections could span any time-skip generated randomly $\mathcal{U}(1,10)$. Backpropagation (BP) through time was run with a learning rate of $\eta = 0.001$ and used Adam momentum with $\mu = 0.9$. Each generated RNNs was trained for 20 BP epochs. Gradient boosting and gradient scaling~\cite{pascanu2013difficulty} were used when the gradient norm was below the threshold of $0.05$ or above the threshold of $1.0$. RNNs were evolved for each of the 30 DJI companies for both 2022 and 2023 DJI company datasets with the RNN that achieved the best validation performance across the 10 repeated EXAMM runs for the stock being selected, resulting in a total of 60 unique RNNs for stock return predictions.

For each DJI company, fully connected two-layer LSTM, GRU, MGU models were also trained using 1000 back-propagation epochs, with Adam as the optimizer and a learning rate of $\eta = 0.0001$. The two-layer memory cell models had two hidden layers of memory cells, fully connected with the number of hidden layers equals to the input layer size. The initial weights for those models were generated randomly using Xavier weight initialization, and the models with highest validation performance over the 10 repeated training runs were selected for testing.

To compare EXAMM's predictive performance with time series forecasting models, we conducted experiments with support vector regression (SVR), random forest regressor (RF), AutoRegressive integrated moving average (ARIMA), and autoregressive recurrent networks (AR). The hyperparameters for those models were tuned using Python's hyperprameter tunning package OPTUNA~\cite{akiba2019optuna} using the validation data. Testing Returns are reported based on the best validation results over 10 repeated training runs. 

\subsection{Transaction Costs}
Transaction costs refer to the costs associated with buying or selling stocks. There are two simple ways to calculate the transaction fees. The first is to use $(Bid-Ask)/2$ as a proxy for transaction costs and apply that transaction cost when selling or buying a stock~\cite{chung2014simple}. When selling a stock, $SellingPrice = StockPrice - (Bid-Ask)/2$, when buying a stock, $BuyingPrice = StockPrice + (Bid-Ask)/2$. The second way to incorporate a transaction fee is to buy the stock with the $Ask$ price and sell with the $Bid$ price. The transaction cost is incorporated in stock transactions when purchasing with a higher stock price (\emph{Ask}) and selling with a lower stock price (\emph{Bid}).

The DJI companies generally have high trading volumes and high liquidity. High liquidity typically results in narrow bid-ask spreads because there are more buyers and sellers. And transaction fees are extremely low compared to its stock price. For example, the average stock price for Apple Inc in 2023 is $\$172.55$, and the average transaction fee calculated by $(Bid-Ask)/2$ is $\$0.08$. 

Table \ref{tab:return_tc} shows the portfolio trading return using the simple long strategy using 2022 and 2023 DJI company datasets. The first column [RET(\%)] shows the portfolio trading return without incorporating any transaction costs. The second column [RET w/ TC1(\%)] shows the portfolio trading return using $(Bid-Ask)/2$ as transaction cost for each trade. The third column [RET w/ TC2(\%)] shows the portfolio return using $Ask$ price to buy and $Bid$ price to sell. From this table we observe that using the same dataset and the same trading strategy, the effects of transaction costs on stock returns are minimal, using either way of incorporating them. Due to this, for our results we do not not incorporate transaction costs in the experiments of the following sections.

\subsection{Return Prediction Performance}
Table \ref{tab:return_all} shows the portfolio returns using the simple long-only strategy in Algorithm \ref{alg:long}.
The trading decisions were made using the different forecasting methods on both the 2022 and 2023 DJI company datasets. Table \ref{tab:market} shows the Buy and Hold return, DJI index return, and S\&P 500 return for 2022 and 2023. The Buy and Hold return is the average return of buying all the stocks at the beginning of the testing period, holding the positions, and selling all stocks at the end of the testing period. An equal amount of capital is used to invest in each stock. The difference between the DJI index return and the buy\& hold return is that DJI is a price weighed index (so they would be purchased in proportion to that initially and then held for the year), but all stocks in the buy\& hold portfolio share equal weight.

Tables \ref{tab:return_all} and \ref{tab:market} show that the portfolio returns using RNNs evolved by EXAMM for predictions are generally higher than the other time series forecasting methods in both 2022 and 2023. In 2022, the DJI index return is $-8.78\%$ and EXAMM predictions incorporating long-only strategy return is $-9.44\%$, while using predictions generated by GRU result in portfolio return of $-8.38\%$, slightly higher than DJI index return and EXAMM predictions. As discussed in Section \ref{long-strategy}, a negative return is inevitable in a bear market. Trading decisions made with any forecasting models will end up having negative investment return using the long-only strategy. However, in 2023, the DJI index return was $13.7\%$ and the S\&P 500 index return was $26.29\%$. Trading decisions made by EXAMM generated RNNs resulted in returns significantly higher than both DJI index and S\&P 500 market index only using DJI's 30 high liquidity companies. Trading decisions made by other forecasting models, such as MGU, GRU, ARIMA, and RF, only generated returns that were higher than DJI index.

\subsection{Portfolio Long-Short Strategy}
Tables \ref{tab:2022_Hybrid} and \ref{tab:2023_hybrid} present the results of employing a simple long-short portfolio trading strategy, which utilizes various stock quantities for daily investments and short selling. This strategy mitigates risk factors across both bull (2023) and bear (2022) market conditions.

In table \ref{tab:2022_Hybrid}, returns higher than the 2022 DJI index return of $-8.78\%$ are highlighted in bold. Given that 2022 was a bear market, the S\&P 500 index exhibited a decline of $-18.11\%$. Thus, returns exceeding this threshold are not distinctly marked, as the majority of the results in the table outperformed the S\&P 500 index. Notably, engaging in more short selling in 2022 yielded returns that not only surpassed market averages but also generated substantial profits in a declining market environment.

Table \ref{tab:2023_hybrid} highlights returns that are higher than the 2023 DJI index return of $-13.70\%$ in bold. Additionally, returns surpassing the 2023 S\&P 500 index of $26.29\%$ are highlighted in green. Optimal results were generally observed when the portfolio allocated between 10\% to one-third of its total size to long positions and short selling, producing profits that consistently outperformed the market indices.

Furthermore, integrating stock return forecasts derived from EXAMM-evolved RNNs with simple long-short trading strategies outperform the market benchmarks in both 2022 and 2023, employing only 30 high liquidity stocks. These results underscore the significant potential of using EXAMM to evolve RNNs for stock return predictions, thereby enhancing the informed decision-making process in portfolio trading for investments.



\section{Conclusion}

This work investigated using neuroevolution to evolve and train time series forecasting recurrent neural networks for stock return predictions in combination with basic stock trading strategies. The RNN models were evolved for each DJI company independently on two datasets prepared for forecasting of the DJI companies' 2022 and 2023 returns, without manually designing or tune the hyperparameters specifically for each model. This significantly reduced the amount of human effort in designing and optimizing these models for each company, and also opens up the opportunity for extending this work to larger portfolios such as the S\&P 500 and S\&P 1500 companies.  

Incorporating simple long-short strategies with the stock return predictions generated by EXAMM evolved RNNs generated significantly higher returns than other time series forecasting models, as well as the DJI index returns. Note that it is more difficult to generate high returns with only 30 high liquidity companies. This is especially for long-short strategies, as the companies are less likely to reduce in value. Further returns could be gained by using custom portfolios which could, for example, add 30 low or medium liquidity companies to allow more potential for short selling generating increased returns.  Further, there is the opportunity to evolve, design and train TSF RNNs which incorporate data from all companies simultaneously, as this could allow them to incorporate correlation between company prices into the forecasts. Finally, returns could potentially be further improved the design of more advanced strategies for buying, selling, shorting and holding stocks, which could be developed using machine learning methods such as reinforcement learning or markov decision processes.

Lastly, the stock market is highly volatile and challenging to accurately forecast, especially when only incorporating time series data. Other external information such as interest rate changes, quarterly business reports and general consumer sentiment can highly influence the performance of a stock. Designing hybrid methods which can incorporate other sources of information could result in better forecasts. Further, we investigated validating our models on a year of data, and tested them without adapting weights or parameters for an additional year, which is an extremely long time for stock forecasting. Refining and updating the models more frequently (\eg~monthly or weekly) or utilizing online forecasting methods could also further improve results.

\bibliographystyle{ACM-Reference-Format}
\bibliography{99-reference}


\begin{thebibliography}{54}


\ifx \showCODEN    \undefined \def \showCODEN     #1{\unskip}     \fi
\ifx \showDOI      \undefined \def \showDOI       #1{#1}\fi
\ifx \showISBNx    \undefined \def \showISBNx     #1{\unskip}     \fi
\ifx \showISBNxiii \undefined \def \showISBNxiii  #1{\unskip}     \fi
\ifx \showISSN     \undefined \def \showISSN      #1{\unskip}     \fi
\ifx \showLCCN     \undefined \def \showLCCN      #1{\unskip}     \fi
\ifx \shownote     \undefined \def \shownote      #1{#1}          \fi
\ifx \showarticletitle \undefined \def \showarticletitle #1{#1}   \fi
\ifx \showURL      \undefined \def \showURL       {\relax}        \fi
\providecommand\bibfield[2]{#2}
\providecommand\bibinfo[2]{#2}
\providecommand\natexlab[1]{#1}
\providecommand\showeprint[2][]{arXiv:#2}

\bibitem[Akiba et~al\mbox{.}(2019)]%
        {akiba2019optuna}
\bibfield{author}{\bibinfo{person}{Takuya Akiba}, \bibinfo{person}{Shotaro Sano}, \bibinfo{person}{Toshihiko Yanase}, \bibinfo{person}{Takeru Ohta}, {and} \bibinfo{person}{Masanori Koyama}.} \bibinfo{year}{2019}\natexlab{}.
\newblock \showarticletitle{Optuna: A next-generation hyperparameter optimization framework}. In \bibinfo{booktitle}{\emph{Proceedings of the 25th ACM SIGKDD international conference on knowledge discovery \& data mining}}. \bibinfo{pages}{2623--2631}.
\newblock


\bibitem[Almahdi and Yang(2017)]%
        {almahdi2017adaptive}
\bibfield{author}{\bibinfo{person}{Saud Almahdi} {and} \bibinfo{person}{Steve~Y Yang}.} \bibinfo{year}{2017}\natexlab{}.
\newblock \showarticletitle{An adaptive portfolio trading system: A risk-return portfolio optimization using recurrent reinforcement learning with expected maximum drawdown}.
\newblock \bibinfo{journal}{\emph{Expert Systems with Applications}}  \bibinfo{volume}{87} (\bibinfo{year}{2017}), \bibinfo{pages}{267--279}.
\newblock


\bibitem[Bali et~al\mbox{.}(2016)]%
        {bali2016empirical}
\bibfield{author}{\bibinfo{person}{Turan~G Bali}, \bibinfo{person}{Robert~F Engle}, {and} \bibinfo{person}{Scott Murray}.} \bibinfo{year}{2016}\natexlab{}.
\newblock \bibinfo{booktitle}{\emph{Empirical asset pricing: The cross section of stock returns}}.
\newblock \bibinfo{publisher}{John Wiley \& Sons}.
\newblock


\bibitem[Belhaj~Slimene and Mamoghli(2019)]%
        {belhaj2019neuroevolution}
\bibfield{author}{\bibinfo{person}{Senda Belhaj~Slimene} {and} \bibinfo{person}{Chokri Mamoghli}.} \bibinfo{year}{2019}\natexlab{}.
\newblock \showarticletitle{NeuroEvolution of Augmenting Topologies for predicting financial distress: A multicriteria decision analysis}.
\newblock \bibinfo{journal}{\emph{Journal of Multi-Criteria Decision Analysis}} \bibinfo{volume}{26}, \bibinfo{number}{5-6} (\bibinfo{year}{2019}), \bibinfo{pages}{320--328}.
\newblock


\bibitem[Cao et~al\mbox{.}(2020)]%
        {cao2020delafo}
\bibfield{author}{\bibinfo{person}{Hieu~K Cao}, \bibinfo{person}{Han~K Cao}, {and} \bibinfo{person}{Binh~T Nguyen}.} \bibinfo{year}{2020}\natexlab{}.
\newblock \showarticletitle{Delafo: An efficient portfolio optimization using deep neural networks}. In \bibinfo{booktitle}{\emph{Advances in Knowledge Discovery and Data Mining: 24th Pacific-Asia Conference, PAKDD 2020, Singapore, May 11--14, 2020, Proceedings, Part I 24}}. Springer, \bibinfo{pages}{623--635}.
\newblock


\bibitem[Cao et~al\mbox{.}(2023)]%
        {cao2023novel}
\bibfield{author}{\bibinfo{person}{Xinwei Cao}, \bibinfo{person}{Adam Francis}, \bibinfo{person}{Xujin Pu}, \bibinfo{person}{Zenan Zhang}, \bibinfo{person}{Vasilios Katsikis}, \bibinfo{person}{Predrag Stanimirovic}, \bibinfo{person}{Ivona Brajevic}, {and} \bibinfo{person}{Shuai Li}.} \bibinfo{year}{2023}\natexlab{}.
\newblock \showarticletitle{A novel recurrent neural network based online portfolio analysis for high frequency trading}.
\newblock \bibinfo{journal}{\emph{Expert Systems with Applications}}  \bibinfo{volume}{233} (\bibinfo{year}{2023}), \bibinfo{pages}{120934}.
\newblock


\bibitem[{Center for Research in Security Prices (CRSP)}(2023)]%
        {crsp_2023}
\bibfield{author}{\bibinfo{person}{{Center for Research in Security Prices (CRSP)}}.} \bibinfo{year}{2023}\natexlab{}.
\newblock \bibinfo{title}{{CRSP US Stock Databases}}.
\newblock \bibinfo{howpublished}{{The University of Chicago Booth School of Business}}.
\newblock
\urldef\tempurl%
\url{http://www.crsp.com}
\showURL{%
\tempurl}
\newblock
\shownote{{Data set}}.


\bibitem[Chen and Zhou(2020)]%
        {chen2020stock}
\bibfield{author}{\bibinfo{person}{Shile Chen} {and} \bibinfo{person}{Changjun Zhou}.} \bibinfo{year}{2020}\natexlab{}.
\newblock \showarticletitle{Stock prediction based on genetic algorithm feature selection and long short-term memory neural network}.
\newblock \bibinfo{journal}{\emph{IEEE Access}}  \bibinfo{volume}{9} (\bibinfo{year}{2020}), \bibinfo{pages}{9066--9072}.
\newblock


\bibitem[Chung and Shin(2018)]%
        {chung2018genetic}
\bibfield{author}{\bibinfo{person}{Hyejung Chung} {and} \bibinfo{person}{Kyung-shik Shin}.} \bibinfo{year}{2018}\natexlab{}.
\newblock \showarticletitle{Genetic algorithm-optimized long short-term memory network for stock market prediction}.
\newblock \bibinfo{journal}{\emph{Sustainability}} \bibinfo{volume}{10}, \bibinfo{number}{10} (\bibinfo{year}{2018}), \bibinfo{pages}{3765}.
\newblock


\bibitem[Chung and Zhang(2014)]%
        {chung2014simple}
\bibfield{author}{\bibinfo{person}{Kee~H Chung} {and} \bibinfo{person}{Hao Zhang}.} \bibinfo{year}{2014}\natexlab{}.
\newblock \showarticletitle{A simple approximation of intraday spreads using daily data}.
\newblock \bibinfo{journal}{\emph{Journal of Financial Markets}}  \bibinfo{volume}{17} (\bibinfo{year}{2014}), \bibinfo{pages}{94--120}.
\newblock


\bibitem[Cipiloglu~Yildiz and Yildiz(2022)]%
        {cipiloglu2022portfolio}
\bibfield{author}{\bibinfo{person}{Zeynep Cipiloglu~Yildiz} {and} \bibinfo{person}{Selim~Baha Yildiz}.} \bibinfo{year}{2022}\natexlab{}.
\newblock \showarticletitle{A portfolio construction framework using {LSTM}-based stock markets forecasting}.
\newblock \bibinfo{journal}{\emph{International Journal of Finance \& Economics}} \bibinfo{volume}{27}, \bibinfo{number}{2} (\bibinfo{year}{2022}), \bibinfo{pages}{2356--2366}.
\newblock


\bibitem[Curtis(2004)]%
        {curtis2004modern}
\bibfield{author}{\bibinfo{person}{Gregory Curtis}.} \bibinfo{year}{2004}\natexlab{}.
\newblock \showarticletitle{Modern portfolio theory and behavioral finance}.
\newblock \bibinfo{journal}{\emph{The Journal of Wealth Management}} \bibinfo{volume}{7}, \bibinfo{number}{2} (\bibinfo{year}{2004}), \bibinfo{pages}{16--22}.
\newblock


\bibitem[Dautel et~al\mbox{.}(2020)]%
        {dautel2020forex}
\bibfield{author}{\bibinfo{person}{Alexander~Jakob Dautel}, \bibinfo{person}{Wolfgang~Karl H{\"a}rdle}, \bibinfo{person}{Stefan Lessmann}, {and} \bibinfo{person}{Hsin-Vonn Seow}.} \bibinfo{year}{2020}\natexlab{}.
\newblock \showarticletitle{Forex exchange rate forecasting using deep recurrent neural networks}.
\newblock \bibinfo{journal}{\emph{Digital Finance}}  \bibinfo{volume}{2} (\bibinfo{year}{2020}), \bibinfo{pages}{69--96}.
\newblock


\bibitem[Elbannan(2015)]%
        {elbannan2015capital}
\bibfield{author}{\bibinfo{person}{Mona~A Elbannan}.} \bibinfo{year}{2015}\natexlab{}.
\newblock \showarticletitle{The capital asset pricing model: an overview of the theory}.
\newblock \bibinfo{journal}{\emph{International Journal of Economics and Finance}} \bibinfo{volume}{7}, \bibinfo{number}{1} (\bibinfo{year}{2015}), \bibinfo{pages}{216--228}.
\newblock


\bibitem[Ferdinand et~al\mbox{.}(2023)]%
        {Ferdinand-2023}
\bibfield{author}{\bibinfo{person}{F.~V. Ferdinand}, \bibinfo{person}{K.~V.~I. Saputra}, \bibinfo{person}{Michelle}, {and} \bibinfo{person}{Johan~Sebastian Edbert}.} \bibinfo{year}{2023}\natexlab{}.
\newblock \showarticletitle{Forecasting Stock Price Index of Four Asian Countries During COVID-19 Pandemic Using ARMA-GARCH and RNN Methods}. In \bibinfo{booktitle}{\emph{2023 IEEE International Conference on Industrial Engineering and Engineering Management (IEEM)}}. \bibinfo{pages}{0974--0978}.
\newblock
\urldef\tempurl%
\url{https://doi.org/10.1109/IEEM58616.2023.10406761}
\showDOI{\tempurl}


\bibitem[Foucault et~al\mbox{.}(2013)]%
        {foucault2013market}
\bibfield{author}{\bibinfo{person}{Thierry Foucault}, \bibinfo{person}{Marco Pagano}, {and} \bibinfo{person}{Ailsa R{\"o}ell}.} \bibinfo{year}{2013}\natexlab{}.
\newblock \bibinfo{booktitle}{\emph{Market liquidity: theory, evidence, and policy}}.
\newblock \bibinfo{publisher}{Oxford University Press, USA}.
\newblock


\bibitem[Gao et~al\mbox{.}(2018)]%
        {gao2018share}
\bibfield{author}{\bibinfo{person}{Shao~En Gao}, \bibinfo{person}{Bo~Sheng Lin}, {and} \bibinfo{person}{Chuin-Mu Wang}.} \bibinfo{year}{2018}\natexlab{}.
\newblock \showarticletitle{Share price trend prediction using CRNN with LSTM structure}. In \bibinfo{booktitle}{\emph{2018 International Symposium on Computer, Consumer and Control (IS3C)}}. IEEE, \bibinfo{pages}{10--13}.
\newblock


\bibitem[Ghosh et~al\mbox{.}(2019)]%
        {ghosh2019stock}
\bibfield{author}{\bibinfo{person}{Achyut Ghosh}, \bibinfo{person}{Soumik Bose}, \bibinfo{person}{Giridhar Maji}, \bibinfo{person}{Narayan Debnath}, {and} \bibinfo{person}{Soumya Sen}.} \bibinfo{year}{2019}\natexlab{}.
\newblock \showarticletitle{Stock price prediction using LSTM on Indian share market}. In \bibinfo{booktitle}{\emph{Proceedings of 32nd international conference on}}, Vol.~\bibinfo{volume}{63}. \bibinfo{pages}{101--110}.
\newblock


\bibitem[Grimes(1977)]%
        {grimes1977stock}
\bibfield{author}{\bibinfo{person}{DH Grimes}.} \bibinfo{year}{1977}\natexlab{}.
\newblock \bibinfo{title}{Stock market logic: a sophisticated approach to profits on Wall Street}.
\newblock
\newblock


\bibitem[Hou et~al\mbox{.}(2020)]%
        {hou2020enriched}
\bibfield{author}{\bibinfo{person}{Xiurui Hou}, \bibinfo{person}{Kai Wang}, \bibinfo{person}{Jie Zhang}, {and} \bibinfo{person}{Zhi Wei}.} \bibinfo{year}{2020}\natexlab{}.
\newblock \showarticletitle{An enriched time-series forecasting framework for long-short portfolio strategy}.
\newblock \bibinfo{journal}{\emph{IEEE Access}}  \bibinfo{volume}{8} (\bibinfo{year}{2020}), \bibinfo{pages}{31992--32002}.
\newblock


\bibitem[Hull(1997)]%
        {hull1997options}
\bibfield{author}{\bibinfo{person}{John~C Hull}.} \bibinfo{year}{1997}\natexlab{}.
\newblock \bibinfo{title}{Options, futures}.
\newblock
\newblock


\bibitem[Iqmal and Putra(2020)]%
        {iqmal2020macroeconomic}
\bibfield{author}{\bibinfo{person}{Fariz~Mohamad Iqmal} {and} \bibinfo{person}{Ivan Gumilar~Sambas Putra}.} \bibinfo{year}{2020}\natexlab{}.
\newblock \showarticletitle{Macroeconomic factors and influence on stock return that impact the corporate values}.
\newblock \bibinfo{journal}{\emph{International Journal of Finance \& Banking Studies (2147-4486)}} \bibinfo{volume}{9}, \bibinfo{number}{1} (\bibinfo{year}{2020}), \bibinfo{pages}{68--75}.
\newblock


\bibitem[Jozefowicz et~al\mbox{.}(2015)]%
        {jozefowicz2015empirical}
\bibfield{author}{\bibinfo{person}{Rafal Jozefowicz}, \bibinfo{person}{Wojciech Zaremba}, {and} \bibinfo{person}{Ilya Sutskever}.} \bibinfo{year}{2015}\natexlab{}.
\newblock \showarticletitle{An empirical exploration of recurrent network architectures}. In \bibinfo{booktitle}{\emph{International Conference on Machine Learning}}. \bibinfo{pages}{2342--2350}.
\newblock


\bibitem[Kumbure et~al\mbox{.}(2022)]%
        {kumbure2022machine}
\bibfield{author}{\bibinfo{person}{Mahinda~Mailagaha Kumbure}, \bibinfo{person}{Christoph Lohrmann}, \bibinfo{person}{Pasi Luukka}, {and} \bibinfo{person}{Jari Porras}.} \bibinfo{year}{2022}\natexlab{}.
\newblock \showarticletitle{Machine learning techniques and data for stock market forecasting: A literature review}.
\newblock \bibinfo{journal}{\emph{Expert Systems with Applications}}  \bibinfo{volume}{197} (\bibinfo{year}{2022}), \bibinfo{pages}{116659}.
\newblock


\bibitem[Leippold et~al\mbox{.}(2022)]%
        {leippold2022machine}
\bibfield{author}{\bibinfo{person}{Markus Leippold}, \bibinfo{person}{Qian Wang}, {and} \bibinfo{person}{Wenyu Zhou}.} \bibinfo{year}{2022}\natexlab{}.
\newblock \showarticletitle{Machine learning in the Chinese stock market}.
\newblock \bibinfo{journal}{\emph{Journal of Financial Economics}} \bibinfo{volume}{145}, \bibinfo{number}{2} (\bibinfo{year}{2022}), \bibinfo{pages}{64--82}.
\newblock


\bibitem[Lemieux et~al\mbox{.}(2014)]%
        {lemieux2014clustering}
\bibfield{author}{\bibinfo{person}{Victoria Lemieux}, \bibinfo{person}{Payam~S Rahmdel}, \bibinfo{person}{Rick Walker}, \bibinfo{person}{BL~William Wong}, {and} \bibinfo{person}{Mark Flood}.} \bibinfo{year}{2014}\natexlab{}.
\newblock \showarticletitle{Clustering techniques and their effect on portfolio formation and risk analysis}. In \bibinfo{booktitle}{\emph{Proceedings of the International Workshop on Data Science for Macro-Modeling}}. \bibinfo{pages}{1--6}.
\newblock


\bibitem[Lyu et~al\mbox{.}(2021a)]%
        {lyu2021experimental}
\bibfield{author}{\bibinfo{person}{Zimeng Lyu}, \bibinfo{person}{AbdElRahman ElSaid}, \bibinfo{person}{Joshua Karns}, \bibinfo{person}{Mohamed Mkaouer}, {and} \bibinfo{person}{Travis Desell}.} \bibinfo{year}{2021}\natexlab{a}.
\newblock \showarticletitle{An Experimental Study of Weight Initialization and Lamarckian Inheritance on Neuroevolution}.
\newblock \bibinfo{journal}{\emph{The 24th International Conference on the Applications of Evolutionary Computation (EvoStar: EvoApps)}} (\bibinfo{year}{2021}).
\newblock


\bibitem[Lyu et~al\mbox{.}(2021b)]%
        {lyu2021improving}
\bibfield{author}{\bibinfo{person}{Zimeng Lyu}, \bibinfo{person}{Joshua Karnas}, \bibinfo{person}{AbdElRahman ElSaid}, \bibinfo{person}{Mohamed Mkaouer}, {and} \bibinfo{person}{Travis Desell}.} \bibinfo{year}{2021}\natexlab{b}.
\newblock \showarticletitle{Improving Distributed Neuroevolution Using Island Extinction and Repopulation}.
\newblock \bibinfo{journal}{\emph{The 24th International Conference on the Applications of Evolutionary Computation (EvoStar: EvoApps)}} (\bibinfo{year}{2021}).
\newblock


\bibitem[Ma et~al\mbox{.}(2021)]%
        {ma2021portfolio}
\bibfield{author}{\bibinfo{person}{Yilin Ma}, \bibinfo{person}{Ruizhu Han}, {and} \bibinfo{person}{Weizhong Wang}.} \bibinfo{year}{2021}\natexlab{}.
\newblock \showarticletitle{Portfolio optimization with return prediction using deep learning and machine learning}.
\newblock \bibinfo{journal}{\emph{Expert Systems with Applications}}  \bibinfo{volume}{165} (\bibinfo{year}{2021}), \bibinfo{pages}{113973}.
\newblock


\bibitem[Malinova and Park(2008)]%
        {malinova2008liquidity}
\bibfield{author}{\bibinfo{person}{Katya Malinova} {and} \bibinfo{person}{Andreas Park}.} \bibinfo{year}{2008}\natexlab{}.
\newblock \showarticletitle{Liquidity, volume, and price behavior: The impact of order vs. quote based trading}. In \bibinfo{booktitle}{\emph{11th Symposium on Finance, Banking, and Insurance, University of Karlsruhe, December}}. Citeseer.
\newblock


\bibitem[McCall(2005)]%
        {mccall2005genetic}
\bibfield{author}{\bibinfo{person}{John McCall}.} \bibinfo{year}{2005}\natexlab{}.
\newblock \showarticletitle{Genetic algorithms for modelling and optimisation}.
\newblock \bibinfo{journal}{\emph{Journal of computational and Applied Mathematics}} \bibinfo{volume}{184}, \bibinfo{number}{1} (\bibinfo{year}{2005}), \bibinfo{pages}{205--222}.
\newblock


\bibitem[Nadkarni and Neves(2018)]%
        {nadkarni2018combining}
\bibfield{author}{\bibinfo{person}{Jo{\~a}o Nadkarni} {and} \bibinfo{person}{Rui~Ferreira Neves}.} \bibinfo{year}{2018}\natexlab{}.
\newblock \showarticletitle{Combining NeuroEvolution and Principal Component Analysis to trade in the financial markets}.
\newblock \bibinfo{journal}{\emph{Expert Systems with Applications}}  \bibinfo{volume}{103} (\bibinfo{year}{2018}), \bibinfo{pages}{184--195}.
\newblock


\bibitem[Ororbia et~al\mbox{.}(2019)]%
        {ororbia2019examm}
\bibfield{author}{\bibinfo{person}{Alexander Ororbia}, \bibinfo{person}{AbdElRahman ElSaid}, {and} \bibinfo{person}{Travis Desell}.} \bibinfo{year}{2019}\natexlab{}.
\newblock \showarticletitle{Investigating Recurrent Neural Network Memory Structures Using Neuro-evolution}. In \bibinfo{booktitle}{\emph{Proceedings of the Genetic and Evolutionary Computation Conference}} (Prague, Czech Republic) \emph{(\bibinfo{series}{GECCO '19})}. \bibinfo{publisher}{ACM}, \bibinfo{address}{New York, NY, USA}, \bibinfo{pages}{446--455}.
\newblock
\showISBNx{978-1-4503-6111-8}
\urldef\tempurl%
\url{https://doi.org/10.1145/3321707.3321795}
\showDOI{\tempurl}


\bibitem[Pahlawan et~al\mbox{.}(2021)]%
        {pahlawan2021stock}
\bibfield{author}{\bibinfo{person}{M~Reza Pahlawan}, \bibinfo{person}{Edwin Riksakomara}, \bibinfo{person}{Raras Tyasnurita}, \bibinfo{person}{Ahmad Muklason}, \bibinfo{person}{Faizal Mahananto}, {and} \bibinfo{person}{Retno~A Vinarti}.} \bibinfo{year}{2021}\natexlab{}.
\newblock \showarticletitle{Stock price forecast of macro-economic factor using recurrent neural network}.
\newblock \bibinfo{journal}{\emph{IAES International Journal of Artificial Intelligence}} \bibinfo{volume}{10}, \bibinfo{number}{1} (\bibinfo{year}{2021}), \bibinfo{pages}{74}.
\newblock


\bibitem[Park et~al\mbox{.}(2020)]%
        {park2020intelligent}
\bibfield{author}{\bibinfo{person}{Hyungjun Park}, \bibinfo{person}{Min~Kyu Sim}, {and} \bibinfo{person}{Dong~Gu Choi}.} \bibinfo{year}{2020}\natexlab{}.
\newblock \showarticletitle{An intelligent financial portfolio trading strategy using deep Q-learning}.
\newblock \bibinfo{journal}{\emph{Expert Systems with Applications}}  \bibinfo{volume}{158} (\bibinfo{year}{2020}), \bibinfo{pages}{113573}.
\newblock


\bibitem[Pascanu et~al\mbox{.}(2013)]%
        {pascanu2013difficulty}
\bibfield{author}{\bibinfo{person}{Razvan Pascanu}, \bibinfo{person}{Tomas Mikolov}, {and} \bibinfo{person}{Yoshua Bengio}.} \bibinfo{year}{2013}\natexlab{}.
\newblock \showarticletitle{On the difficulty of training recurrent neural networks}. In \bibinfo{booktitle}{\emph{International Conference on Machine Learning}}. \bibinfo{pages}{1310--1318}.
\newblock


\bibitem[Pawar et~al\mbox{.}(2019)]%
        {pawar2019stock}
\bibfield{author}{\bibinfo{person}{Kriti Pawar}, \bibinfo{person}{Raj~Srujan Jalem}, {and} \bibinfo{person}{Vivek Tiwari}.} \bibinfo{year}{2019}\natexlab{}.
\newblock \showarticletitle{Stock market price prediction using LSTM RNN}. In \bibinfo{booktitle}{\emph{Emerging Trends in Expert Applications and Security: Proceedings of ICETEAS 2018}}. Springer, \bibinfo{pages}{493--503}.
\newblock


\bibitem[Pellon~Consunji(2021)]%
        {pellon2021evotrader}
\bibfield{author}{\bibinfo{person}{Martin Pellon~Consunji}.} \bibinfo{year}{2021}\natexlab{}.
\newblock \showarticletitle{EvoTrader: Automated Bitcoin Trading Using Neuroevolutionary Algorithms on Technical Analysis and Social Sentiment Data}. In \bibinfo{booktitle}{\emph{Proceedings of the 2021 4th International Conference on Algorithms, Computing and Artificial Intelligence}}. \bibinfo{pages}{1--9}.
\newblock


\bibitem[Ratanapakorn and Sharma(2007)]%
        {ratanapakorn2007dynamic}
\bibfield{author}{\bibinfo{person}{Orawan Ratanapakorn} {and} \bibinfo{person}{Subhash~C Sharma}.} \bibinfo{year}{2007}\natexlab{}.
\newblock \showarticletitle{Dynamic analysis between the US stock returns and the macroeconomic variables}.
\newblock \bibinfo{journal}{\emph{Applied Financial Economics}} \bibinfo{volume}{17}, \bibinfo{number}{5} (\bibinfo{year}{2007}), \bibinfo{pages}{369--377}.
\newblock


\bibitem[Reddy(2019)]%
        {reddy2019predicting}
\bibfield{author}{\bibinfo{person}{C~Viswanatha Reddy}.} \bibinfo{year}{2019}\natexlab{}.
\newblock \showarticletitle{Predicting the stock market index using stochastic time series ARIMA modelling: The sample of BSE and NSE}.
\newblock \bibinfo{journal}{\emph{Indian Journal of Finance}} \bibinfo{volume}{13}, \bibinfo{number}{8} (\bibinfo{year}{2019}), \bibinfo{pages}{7--25}.
\newblock


\bibitem[Selvin et~al\mbox{.}(2017)]%
        {sreelekshmy-2017}
\bibfield{author}{\bibinfo{person}{Sreelekshmy Selvin}, \bibinfo{person}{R Vinayakumar}, \bibinfo{person}{E.~A Gopalakrishnan}, \bibinfo{person}{Vijay~Krishna Menon}, {and} \bibinfo{person}{K.~P. Soman}.} \bibinfo{year}{2017}\natexlab{}.
\newblock \showarticletitle{Stock price prediction using LSTM, RNN and CNN-sliding window model}. In \bibinfo{booktitle}{\emph{2017 International Conference on Advances in Computing, Communications and Informatics (ICACCI)}}. \bibinfo{pages}{1643--1647}.
\newblock
\urldef\tempurl%
\url{https://doi.org/10.1109/ICACCI.2017.8126078}
\showDOI{\tempurl}


\bibitem[Sen et~al\mbox{.}(2021)]%
        {sen2021stock}
\bibfield{author}{\bibinfo{person}{Jaydip Sen}, \bibinfo{person}{Abhishek Dutta}, {and} \bibinfo{person}{Sidra Mehtab}.} \bibinfo{year}{2021}\natexlab{}.
\newblock \showarticletitle{Stock portfolio optimization using a deep learning LSTM model}. In \bibinfo{booktitle}{\emph{2021 IEEE Mysore sub section international conference (MysuruCon)}}. IEEE, \bibinfo{pages}{263--271}.
\newblock


\bibitem[Silva et~al\mbox{.}(2020)]%
        {silva2020automated}
\bibfield{author}{\bibinfo{person}{Thalita~R Silva}, \bibinfo{person}{Audeliano~W Li}, {and} \bibinfo{person}{Edson~O Pamplona}.} \bibinfo{year}{2020}\natexlab{}.
\newblock \showarticletitle{Automated trading system for stock index using LSTM neural networks and risk management}. In \bibinfo{booktitle}{\emph{2020 international joint conference on neural networks (IJCNN)}}. IEEE, \bibinfo{pages}{1--8}.
\newblock


\bibitem[Smith et~al\mbox{.}(2024)]%
        {smith2024stock}
\bibfield{author}{\bibinfo{person}{Nathan Smith}, \bibinfo{person}{Vivek Varadharajan}, \bibinfo{person}{Dinesh Kalla}, \bibinfo{person}{Ganesh~R Kumar}, {and} \bibinfo{person}{Fnu Samaah}.} \bibinfo{year}{2024}\natexlab{}.
\newblock \showarticletitle{Stock Closing Price and Trend Prediction with LSTM-RNN}.
\newblock \bibinfo{journal}{\emph{Journal of Artificial Intelligence and Big Data}} (\bibinfo{year}{2024}), \bibinfo{pages}{1--13}.
\newblock


\bibitem[Stulz(2008)]%
        {stulz2008rethinking}
\bibfield{author}{\bibinfo{person}{Ren{\'e}~M Stulz}.} \bibinfo{year}{2008}\natexlab{}.
\newblock \showarticletitle{Rethinking risk management}.
\newblock In \bibinfo{booktitle}{\emph{Corporate Risk Management}}. \bibinfo{publisher}{Columbia University Press}, \bibinfo{pages}{87--120}.
\newblock


\bibitem[Ta et~al\mbox{.}(2020)]%
        {ta2020portfolio}
\bibfield{author}{\bibinfo{person}{Van-Dai Ta}, \bibinfo{person}{Chuan-Ming Liu}, {and} \bibinfo{person}{Direselign~Addis Tadesse}.} \bibinfo{year}{2020}\natexlab{}.
\newblock \showarticletitle{Portfolio optimization-based stock prediction using long-short term memory network in quantitative trading}.
\newblock \bibinfo{journal}{\emph{Applied Sciences}} \bibinfo{volume}{10}, \bibinfo{number}{2} (\bibinfo{year}{2020}), \bibinfo{pages}{437}.
\newblock


\bibitem[Thakkar and Chaudhari(2022)]%
        {thakkar2022information}
\bibfield{author}{\bibinfo{person}{Ankit Thakkar} {and} \bibinfo{person}{Kinjal Chaudhari}.} \bibinfo{year}{2022}\natexlab{}.
\newblock \showarticletitle{Information fusion-based genetic algorithm with long short-term memory for stock price and trend prediction}.
\newblock \bibinfo{journal}{\emph{Applied Soft Computing}}  \bibinfo{volume}{128} (\bibinfo{year}{2022}), \bibinfo{pages}{109428}.
\newblock


\bibitem[Tsay(2005)]%
        {tsay2005analysis}
\bibfield{author}{\bibinfo{person}{Ruey~S Tsay}.} \bibinfo{year}{2005}\natexlab{}.
\newblock \bibinfo{booktitle}{\emph{Analysis of financial time series}}.
\newblock \bibinfo{publisher}{John wiley \& sons}.
\newblock


\bibitem[Wang et~al\mbox{.}(2017)]%
        {wang2017session}
\bibfield{author}{\bibinfo{person}{Shuhao Wang}, \bibinfo{person}{Cancheng Liu}, \bibinfo{person}{Xiang Gao}, \bibinfo{person}{Hongtao Qu}, {and} \bibinfo{person}{Wei Xu}.} \bibinfo{year}{2017}\natexlab{}.
\newblock \showarticletitle{Session-based fraud detection in online e-commerce transactions using recurrent neural networks}. In \bibinfo{booktitle}{\emph{Machine Learning and Knowledge Discovery in Databases: European Conference, ECML PKDD 2017, Skopje, Macedonia, September 18--22, 2017, Proceedings, Part III 10}}. Springer, \bibinfo{pages}{241--252}.
\newblock


\bibitem[Xie et~al\mbox{.}(2023)]%
        {xie2023efficient}
\bibfield{author}{\bibinfo{person}{Xiangning Xie}, \bibinfo{person}{Xiaotian Song}, \bibinfo{person}{Zeqiong Lv}, \bibinfo{person}{Gary~G Yen}, \bibinfo{person}{Weiping Ding}, {and} \bibinfo{person}{Yanan Sun}.} \bibinfo{year}{2023}\natexlab{}.
\newblock \showarticletitle{Efficient evaluation methods for neural architecture search: A survey}.
\newblock \bibinfo{journal}{\emph{arXiv preprint arXiv:2301.05919}} (\bibinfo{year}{2023}).
\newblock


\bibitem[Yao et~al\mbox{.}(2018)]%
        {yao2018high}
\bibfield{author}{\bibinfo{person}{Siyu Yao}, \bibinfo{person}{Linkai Luo}, {and} \bibinfo{person}{Hong Peng}.} \bibinfo{year}{2018}\natexlab{}.
\newblock \showarticletitle{High-frequency stock trend forecast using LSTM model}. In \bibinfo{booktitle}{\emph{2018 13th International Conference on Computer Science \& Education (ICCSE)}}. IEEE, \bibinfo{pages}{1--4}.
\newblock


\bibitem[Zhang and Huang(2021)]%
        {zhang2021option}
\bibfield{author}{\bibinfo{person}{Junhuan Zhang} {and} \bibinfo{person}{Wenjun Huang}.} \bibinfo{year}{2021}\natexlab{}.
\newblock \showarticletitle{Option hedging using LSTM-RNN: an empirical analysis}.
\newblock \bibinfo{journal}{\emph{Quantitative Finance}} \bibinfo{volume}{21}, \bibinfo{number}{10} (\bibinfo{year}{2021}), \bibinfo{pages}{1753--1772}.
\newblock


\bibitem[Zhang et~al\mbox{.}(2021)]%
        {zhang2021predicting}
\bibfield{author}{\bibinfo{person}{Xiaoci Zhang}, \bibinfo{person}{Naijie Gu}, \bibinfo{person}{Jie Chang}, {and} \bibinfo{person}{Hong Ye}.} \bibinfo{year}{2021}\natexlab{}.
\newblock \showarticletitle{Predicting stock price movement using a DBN-RNN}.
\newblock \bibinfo{journal}{\emph{Applied Artificial Intelligence}} \bibinfo{volume}{35}, \bibinfo{number}{12} (\bibinfo{year}{2021}), \bibinfo{pages}{876--892}.
\newblock


\bibitem[Zhao et~al\mbox{.}(2021)]%
        {zhao2021prediction}
\bibfield{author}{\bibinfo{person}{Jinghua Zhao}, \bibinfo{person}{Dalin Zeng}, \bibinfo{person}{Shuang Liang}, \bibinfo{person}{Huilin Kang}, {and} \bibinfo{person}{Qinming Liu}.} \bibinfo{year}{2021}\natexlab{}.
\newblock \showarticletitle{Prediction model for stock price trend based on recurrent neural network}.
\newblock \bibinfo{journal}{\emph{Journal of Ambient Intelligence and Humanized Computing}}  \bibinfo{volume}{12} (\bibinfo{year}{2021}), \bibinfo{pages}{745--753}.
\newblock


\end{thebibliography}


\end{document}